\documentclass[prb,aps,superscriptaddress,eqsecnum,floatfix]{revtex4-1}
\expandafter\let\csname equation*\endcsname\relax

\expandafter\let\csname endequation*\endcsname\relax
\usepackage{amsmath}
\usepackage{amssymb}
\usepackage{bm}

\usepackage{empheq}
\usepackage{dsfont}
\usepackage{graphicx}
\usepackage{subfig}
\usepackage{hyperref}
\usepackage{stackrel}
\usepackage{color,epsfig}
\usepackage{comment}

\usepackage{colortbl}

\input{epsf}

\numberwithin{equation}{section}

\newcommand{\BS}{\boldsymbol \sigma}
\newcommand{\Bgrad}{\boldsymbol \nabla}
\newcommand{\sign}{{\rm\,  sgn}}

\newcommand{\bb}{\begin{equation}}
\newcommand{\ee}{\end{equation}}
\newcommand{\eqb}{\begin{eqnarray}}
\newcommand{\eqf}{\end{eqnarray}}

\def \ee{\end{equation}}
\def \be{\begin{equation}}
\def \eea{\end{eqnarray}}
\def \bea{\begin{eqnarray}}

\begin{document}

\title{Optical Conductivity in an effective model for Graphene: Finite temperature corrections.}

\author{Horacio Falomir}
\affiliation{IFLP, CONICET - Departamento de F\'{\i}sica, Fac.\ de Ciencias Exactas de la UNLP, C.C. 67, (1900) La Plata, Argentina.}
\email{falomir@fisica.unlp.edu.ar}
\author{Enrique Mu\~noz}
\affiliation{Instituto de F\'isica, Pontificia Universidad Cat\'olica de Chile, 
Avenida Vicu\~na Mackenna 4860, Santiago, Chile.}
\affiliation{Center for Nanotechnology and Advanced Materials CIEN-UC, Avenida Vicu\~na Mackenna 4860, Santiago, Chile.}
\email{munozt@fis.puc.cl}
\author{Marcelo Loewe}
\affiliation{Instituto de F\'isica, Pontificia Universidad Cat\'olica de Chile, 
Avenida Vicu\~na Mackenna 4860, Santiago, Chile.}
\affiliation{Centre for Theoretical and Mathematical Physics, University of Cape Town, Rondebosch 770, South Africa.}
\affiliation{Centro Cient\'ifico Tecnol\'ogico de Valpara\'iso, CCTVAL, Universidad T\'ecnica Federico Santa Mar\'ia, Casilla 110-V, Valpara\'iso, Chile.}
\email{mloewe@fis.puc.cl}
\author{Renato Zamora}
\affiliation{Instituto de Ciencias B\'asicas, Universidad Diego Portales, Casilla 298-V, Santiago, Chile.}
\affiliation{Centro de Investigaci\'on y Desarrollo de Ciencias Aeroespaciales (CIDCA), Fuerza A\'erea de Chile,
Santiago 8020744, Chile}
\email{rzamorajofre@gmail.com}

\begin{abstract}
In this article, we
investigate the temperature and chemical potential dependence of the optical conductivity of graphene, within a field theoretical representation in the continuum approximation,
arising from an underlying tight-binding atomistic model, that includes up to next-to-nearest
neighbor coupling. Our calculations allow us to obtain the dependence of the optical conductivity on frequency, temperature and finite chemical potential, generalizing
our previouly reported calculations at zero temperature, and
reproducing the universal and experimentally verified value at zero frequency. 
\end{abstract}


\maketitle
\section{Introduction}
Graphene, a monolayer of carbon atoms arranged in a honeycomb lattice with $C_{3v}\otimes Z_2$ symmetry \cite{Wallace_47}, possesses an electronic spectrum that displays two non-equivalent points $K_{+},\,K_{-}$ where the conduction and valence bands touch, and in whose vicinity
the dispersion relation is approximately linear. The electronic spectrum is correctly described by an atomistic tight-binding model that, when including up to first nearest-neighbors coupling, leads to an effective low-energy continuum model describing relativistic Dirac fermions in 2D.
This minimal tight-binding model can be extended upon including second nearest-neighbors couplings, that in the continuum representation leads to an effective field theory with a quadratic contribution to the linear Dirac dispersion~\cite{GNAQ}. Transparency is a physical property
determined by the optical conductivity, i.e. the linear response to an {external} electromagnetic field.
Several experiments confirm~\cite{Nair,WASSEI201052,Ma2013,Mak24082010,shou,FV-2012,mariel,FV-2016,FV-2011B,FV-2011,Fial-2011,david,saul,Merthe} that the measured transmittance is indeed compatible with the effective single-particle model of relativistic Dirac fermions in
graphene, as supported from a number of theoretical works~\cite{FV-2012,mariel,FV-2016,FV-2011B,FV-2011,Fial-2011,david,saul,Merthe}. Among several physical effects that may induce deviations from the single-particle Dirac dispersion continuum model, such as electron-electron Coulomb interactions\cite{Kotov_12,DasSarma_11}, lattice phonons\cite{Hwang_2008,Tse_07,Munoz_012,Munoz_16,Kubakaddi_09},
impurities \cite{Ando_2006,CastroNeto_09,Peres_10} and different forms of quenched disorder\cite{DasSarma_11,CastroNeto_09}, we shall focus on the
contribution to the optical conductivity that arises from the next-to-nearest neighbors coupling in the atomistic Hamiltonian,
included as a quadratic correction to the kinetic energy operator within a continuum effective model for graphene \cite{Cond_T0}. Such a model has been considered by some of us in Ref.~\cite{GNAQ} to fully account for the Anomalous Integer Quantum Hall Effect in this material and the underlying wave equation is referred to in literature as Second Order  Dirac Equation~\cite{second}.
Notice that this is an isotropic model in which, the quadratic (anisotropic next to leading) term in the dispersion relation coming from the nearest neighbor sites has been shown to give a vanishing contribution to the Hamiltonian spectrum at first order in perturbation theory, thus justifying the consideration of the quadratic (isotropic) leading contribution of next-to-nearest neighbors in the honeycomb array~\cite{GNAQ}.
In a previous article\cite{Cond_T0}, we investigated the frequency dependence of the zero-temperature optical conductivity of graphene, calculated in the Kubo linear response approximation \cite{Kubo_II,Wen,Stefanucci}, when these next-to-nearest neighbors corrections are included in an effective field theory on the closed time path (CTP) (or Keldysh ~\cite{Rammer,Stefanucci}) formalism.
In our present article, we extend this analysis to include finite temperature and finite chemical potential effects.

Along the previously exposed ideas, we have organized the remaining of this article as follows: In Sect.~\ref{Ec-Mov}, we present the details of the model. In Sect.~\ref{vacpol} we present the Matsubara formalism to calculate the vacuum polarization tensor in the Euclidean representation, to finally obtain the optical conductivity from the vacuum polarization tensor
via analytic continuation to real frequency space. We discuss our findings in Sect.~\ref{conclusions}. Some calculation details are presented in two Appendices.

\section{Lagrangian, conserved current and generating functional}\label{Ec-Mov}

\begin{figure}[tbp]
\centering
\epsfig{file=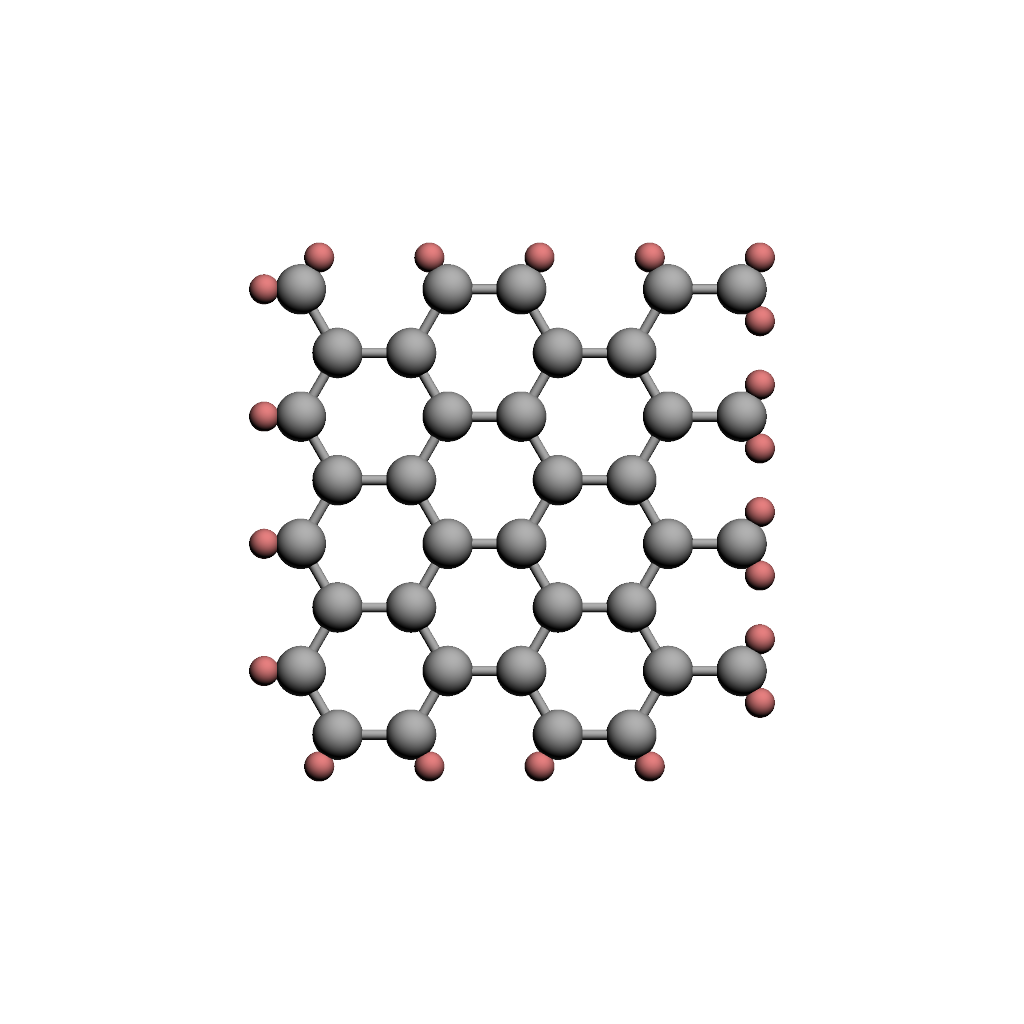,width=0.5\columnwidth}
\caption{(Color online) Sketch of the crystal structure of graphene. The honeycomb array is described in terms of two overlapping triangular sublattices. 
}
\label{fig1a}
\end{figure}

Graphene crystal structure, as sketched in Fig.~\ref{fig1a}, is described in terms of two overlapping triangular (Bravais) sublattices. The band structure obtained from an atomistic, tight-binding description including up to the next-to-nearest neighbors contribution is of the form
\begin{equation}
E_\pm(\mathbf{k})=\pm t\sqrt{f(\mathbf{k})}-t'[f(\mathbf{k})-3],
\label{eq_dispersion_1}
\end{equation}
where $t$ and $t'$ are the nearest and next-to-nearest hopping parameters and
\begin{equation}
f(\mathbf{k})=3+4\cos\left( \frac{3k_x a}{2}\right)\cos\left( \frac{\sqrt{3}k_y a}{2}\right)+2\cos(\sqrt{3}k_ya)\;.
\label{eq_dispersion_2}
\end{equation}
Here, $a\simeq 1.42\AA$ is the interatomic distance\cite{CastroNeto_09}. The literature reports a value\cite{CastroNeto_09} $t \sim 2.8\,eV$, while for the second nearest-neighbour coupling the 
reference values are not so precisely established, but reported in the range\cite{CastroNeto_09} $0.02 t \le t' \le 0.2t$.

The points $K_+$ and $K_-$ at which $f(K_\pm)=0$ define the so-called Dirac points. Around $K_+$,
\begin{equation}
E_\pm(\mathbf{k}+K_+)=\pm t\left[\frac{3}{2}a |\mathbf{k}|-\frac{3}{8}a^2 \mathbf{k}^2 \sin(3\vartheta) \right]+t'\left[-\frac{9}{4}a^3\mathbf{k}^2+3 \right]+{\cal O}(|\mathbf{k}|^3),
\label{nnn}
\end{equation}
with $\tan(\vartheta)=k_y/k_x$. Around the $K_-$ point, one just needs to replace $\vartheta\to -\vartheta$ in Eq.~(\ref{nnn}). The isotropic portion of the model in Eq.~(\ref{nnn}) was first considered in Ref.~\cite{GNAQ} as a natural framework to explain the Anomalous Integer Quantum Hall Effect in graphene. Moreover, as previously mentioned, the anisotropic quadratic term, so called trigonal warping, in this effective dispersion relation was shown not to contribute to the energy spectrum at first order in perturbation theory~\cite{GNAQ}, thus justifying to retain just the isotropic terms up to this order in the pseudo-momenta.

In the presence of electromagnetic interactions, the model in the continuum approximation is described by an effective field theory with 
the Lagrangian~\cite{GNAQ,Cond_T0}
\begin{eqnarray}
      \mathcal{L}&:=&\frac{i}{2}\left[\psi^\dagger \, \partial_t \psi - \partial_t\psi^\dagger\, \psi\right]+
      \psi^\dagger e A_0  \psi \nonumber\\
  &&-\frac{1}{2m} \left\{ \left[\left(\mathbf{p}-e\mathbf{A}+\theta\BS\right)\psi\right]^\dagger \cdot \left[\left(\mathbf{p}-e\mathbf{A}+\theta\BS\right)\psi\right]-2\theta^2\psi^\dagger\psi\right\}\nonumber\\
   &=& \frac{i}{2}\left[\psi^\dagger \, \partial_t \psi - \partial_t\psi^\dagger\, \psi\right]
   -\frac{1}{2m} \left\{\Bgrad\psi^\dagger \cdot \Bgrad \psi +
   i \Bgrad\psi^\dagger \cdot  \left(-e\mathbf{A}+\theta\BS\right)\psi -\right.\nonumber\\
   &&\left. -i \psi^\dagger  \left(-e\mathbf{A}+\theta\BS\right) \cdot\Bgrad\psi+
   \psi^\dagger  \left[\left(-e\mathbf{A}+\theta\BS\right)^2- 2 \theta^2 \right]\psi
   \right\}
\,,\label{L1}
\end{eqnarray}
where $\theta=m v_F$ and $m = \pm 2 \hbar^2/(9 t' a^2)$, where the sign depends on each Dirac cone $K_{\pm}$. A summary of the numerical values for the relevant parameters 
of the model is presented in Table \ref{tab:table1}.

\begin{table}[h!]
  \begin{center}
    \caption{Parameters of the model}
    \label{tab:table1}
    \begin{tabular}{|l|c|r} 
    \hline
      $a\, (\AA)$\cite{CastroNeto_09} & $1.42$\\
      $t\, (eV)$\cite{CastroNeto_09} & $2.8$\\
      $t'\, (eV)$\cite{CastroNeto_09} & $\sim 0.056 - 0.56$\\
      $m\,(kg)$ & $1.37\times 10^{-29} - 1.37\times 10^{-30}$\\
      $v_f\,(m/s)$\cite{CastroNeto_09} & $\sim 10^6$\\
      $m v_f^2\,(eV)$ & $7 - 70$\\
      \hline
    \end{tabular}
  \end{center}
\end{table}

Here, the 3-momentum is $p^{\mu}=(p^0,\mathbf{p})$, with $\mathbf{p} = (p^1,p^2)$.
The vector potential $\mathbf{A} = (A^{1},A^2)$, whereas $\bm{\sigma} = (\sigma^1,\sigma^2)$ are Pauli matrices.
In this model, $\psi^\dagger$ and $\psi$ are regarded as independent fields whose equations of motion are derived from the variation of the action,
\begin{eqnarray}
        \frac{\partial \mathcal{L}}{\partial \psi^\dagger}-\partial_t \left( \frac{\partial\mathcal{L}}{\partial \left(\partial_t \psi^\dagger\right)}\right) -
    \Bgrad\cdot \left( \frac{\partial\mathcal{L}}{\partial \left(\Bgrad \psi^\dagger\right)}\right)\nonumber\\
   &&\hspace{-5cm}   =i \partial_t \psi- \frac{1}{2m} \left[\left(\mathbf{p}-e\mathbf{A}+\theta\BS\right)^2
      -2\theta^2\right] \psi =0\,,\label{L2}
\end{eqnarray}
and similarly for $\psi$.

\medskip

N{\oe}ther's Theorem leads to the existence of a locally conserved current, 
whose time-component defines the local charge density \cite{Cond_T0}
\begin{equation}\label{L6}
    j^0 = e \, \psi^\dagger \psi,
\end{equation}
while the spatial components define the current density \cite{Cond_T0}
\begin{equation}\label{L7}
    {j^k}=\frac{e}{2m}\left\{
    i \left(\partial^k \psi^\dagger\,  \psi - \psi^\dagger \, \partial^k\psi \right)
    + 2 \psi^\dagger\left(-e {A^k}+\theta\sigma^k\right)\psi\right\}\,.
\end{equation}
It is straightforward to verify, from  the equations of motion, that $j^{\mu}$ is conserved \cite{Cond_T0},
\begin{equation}\label{L8}
    \partial_\mu j^\mu=\partial_t j^0 - \Bgrad \cdot \mathbf{j}=0\,.
\end{equation}
Notice also that we can write \cite{Cond_T0}
\begin{equation}\label{JderivGamma}
    j^\mu(x)= \frac{\delta}{\delta A_\mu(x)} \int \mathcal{L}(y) \,  d^3 y\,.
\end{equation}

In our previous work\cite{Cond_T0}, we developed a generating functional on the CTP (or Keldysh contour) for the effective field theory in Eq.(\ref{nnn}), defined
as
\begin{eqnarray}
{Z_{\gamma}[A]} = \int\mathcal{D}\psi^{\dagger}(\mathbf{x},\tau) \mathcal{D}\psi(\mathbf{x},\tau) e^{i \int_{\gamma} d\tau \int d^{2}\mathbf{x} \mathcal{L}[\psi^{\dagger}(\mathbf{x},\tau),\psi(\mathbf{x},\tau) ]},
\label{eq_CTP_gen}
\end{eqnarray}
with $\gamma = \gamma_{-}\oplus\gamma_{+}$, such that $\gamma_{-}$ represents the time-ordered branch of the contour, while $\gamma_{+}$ is the anti-time-ordered branch (see Ref.~\cite{Cond_T0} for details).
From the CTP functional defined in Eq.~(\ref{eq_CTP_gen}), we generate the average current components as follows\cite{Cond_T0} 
\begin{eqnarray}
    -i  \frac{\delta\log Z_{\gamma}[A]}{\delta A_\mu(x)}&=&
    \frac{1}{Z_{\gamma}[A]}  \int \mathcal{D} \psi^\dagger \mathcal{D}\psi \, e^{\displaystyle
    i \int_{\gamma} d^3 y \mathcal{L}(y)} j^\mu(x)\nonumber\\
      &=&   \left\langle j^\mu(x) \right\rangle,\label{L10}
\end{eqnarray}
while the second functional derivative gives the current-current correlation function\cite{Cond_T0},
\begin{eqnarray}
     (-i )^2  \frac{\delta^2\log Z_{\gamma}[A]}{\delta A_\mu(x) \delta A_\nu(y)}   =-i  \left\langle \frac{\delta j^\mu(x)}{\delta A_\nu(y)} \right\rangle +
   \left\langle \mathcal{T} j^\mu(x) j^\nu(y) \right\rangle -
   \left\langle j^\mu(x)  \right\rangle \left\langle j^\nu(y) \right\rangle\,.\label{correl-js}
\end{eqnarray}
Here, the first term is the \emph{diamagnetic contribution} \cite{Altland-Simons,Cond_T0}
\begin{equation}\label{diamagnetic-term}
   \begin{array}{c} \displaystyle
      \left\langle \frac{\delta j^\mu(x)}{\delta A_\nu(y)} \right\rangle=
      \delta^{\mu k} \delta^{\nu}_k \left( - \frac{{e^2}}{m^2}\right) \left\langle \psi^\dagger(x) \psi(x)\right\rangle \delta^{(3)} \left( x-y \right),
   \end{array}
\end{equation}
and the others are the \emph{paramagnetic} ones.

The currents are defined in normal order with respect to the fermionic field, so that $\left.\left\langle j^\mu(x) \right\rangle\right|_{A=0}=0$. The \emph{\emph{linear response}} of the system to the external electromagnetic field is described by the second derivative in Eq.~(\ref{correl-js}) evaluated at $A_\mu=0$~\cite{Altland-Simons,Cond_T0},
\begin{eqnarray}
      K^{\mu\nu}(x,y) &=& \left. (-i )^2  \frac{\delta^2\log Z_{\gamma}[A]}{\delta A_\mu(x) \delta A_\nu(y)} \right|_{A=0}
			= K^{\nu\mu}(y,x) \nonumber\\
     &=&\left\langle \mathcal{T} j^\mu(x) j^\nu(y) \right\rangle_0\,. \label{Kmunu}
\end{eqnarray}

The spatial components of the current are given by\cite{Cond_T0}
\begin{eqnarray}
j^{k}(x)\Bigg|_{A=0} &=& \frac{e}{2m}\left\{
i\partial^{k}\psi^{\dagger}(x) \psi(x) - i\psi^{\dagger}(x) \partial^{k}\psi(x) + 2\theta \psi^{\dagger}(x)\sigma^{k}\psi(x)
\right\}\nonumber\\
&\equiv& \psi^{\dagger}_a(x)\hat{D}_{ab}^{k}\psi_b(x)\;,
\end{eqnarray}

where we have defined the differential operators\cite{Cond_T0}
\begin{eqnarray}
\hat{D}_{ab}^{k} = \frac{e}{2m}\left\{-i\overleftrightarrow{\partial}^{k} \delta_{ab}+ 2\theta \left[\sigma^k\right]_{ab}
\right\}\;.
\end{eqnarray}

Applying Wick's theorem \cite{Rammer,Kamenev_011,Stefanucci} on the CTP for the definition of the current-correlator (correlators associated to disconnected diagrams vanish), we obtain~\cite{Cond_T0}:
\begin{eqnarray}
 \langle \mathcal{T} j^{k}(x) j^{l} (y)\rangle
&=& \langle  \mathcal{T} \psi_{a}^{\dagger}(x)\hat{D}_{ab}^k \psi_{b}(x) \psi_c^{\dagger}(y)\hat{D}_{cd}^{l}\psi_d(y)\rangle\nonumber\\
&=& -\hat{D}_{ab}^k \hat{D}_{cd}^{l}
  \langle  \mathcal{T} \psi_{b}(x) \psi_c^{\dagger}(y)\rangle \langle  \mathcal{T} \psi_{d}(y) \psi_a^{\dagger}(x)\rangle\;.
\end{eqnarray}
The previous relation allows us to define the corresponding components of the polarization tensor in the CTP contour indices $\alpha,\beta = \pm$,
\begin{eqnarray}
K^{k l}_{\alpha\beta}(x,y) &=& \langle  \mathcal{T} j^{k}_{\alpha}(x) j^{l}_{\beta} (y)\rangle\nonumber\\
&=& -\hat{D}_{ab}^k \hat{D}_{cd}^{l} \Delta_{bc}^{\alpha\beta}(x,y)\Delta_{da}^{\beta\alpha}(y,x)\;.
\end{eqnarray}
As discussed in detail in Ref.~\cite{Cond_T0}, the retarded component of the polarization tensor is obtained from the combination  
\begin{eqnarray}
K_{R}^{kl}(x,y) &=& K_{--}^{kl}(x,y) - K_{-+}^{kl}(x,y)\nonumber\\
&=& \hat{D}_{ab}^k \hat{D}_{cd}^{l} \Bigg\{
\Delta_{bc}^{F}(x,y) \Delta_{da}^{A}(y,x) + \Delta_{bc}^{R}(x,y) \Delta_{da}^{F}(y,x) \nonumber\\
&&- \Delta_{bc}^{R}(x,y) \Delta_{da}^{A}(y,x)
\Bigg\}\;.
\label{eq_KR}
\end{eqnarray}
In terms of Fourier transforms, 
\begin{equation}\label{psi-de-p}
        \psi(x)= \frac{1}{\left( 2 \pi\right)^{3/2}} \int d^3p \, e^{-i p\cdot x} \tilde{\psi}(p)\,,
  \qquad
      \psi^\dagger(x)= \frac{1}{\left( 2 \pi\right)^{3/2}} \int d^3p \, e^{i p\cdot x} \tilde{\psi}^\dagger(p)\,,
    \end{equation}
we have~\cite{Cond_T0}
\begin{eqnarray}
\Delta_{ab}^{\alpha\beta}(x,y) \equiv \Delta_{ab}^{\alpha\beta}(x-y) = \int\frac{d^3 p }{(2\pi)^3} e^{i(x-y)\cdot p}\tilde{\Delta}_{ab}^{\alpha\beta}(p).
\end{eqnarray}
Here, the different propagators for the Hamiltonian model considered are, in Fourier space (F: Feynman, R: Retarded, A: Advanced),
\begin{eqnarray}
\tilde{\Delta}^{F}(p) &=& \tilde{\Delta}_{--}(p)= i\frac{p_0 - \frac{\mathbf{p}^2}{2m} + v_F \mathbf{p}\cdot\bm{\sigma}}{\left( p_0 - \frac{\mathbf{p}^2}{2m}\right)^2 - v_F^2\mathbf{p}^2 + i\epsilon'}\nonumber\\
&&= i\frac{p_0 - \frac{\mathbf{p}^2}{2m} + v_F \mathbf{p}\cdot\bm{\sigma}}{\left( p_0 + i\epsilon - \frac{\mathbf{p}^2}{2m} - v_F|\mathbf{p}|\right)\left( p_0 - i\epsilon - \frac{\mathbf{p}^2}{2m} + v_F|\mathbf{p}|\right)}\;,
\\
\tilde{\Delta}^{R}(p) &=& i\frac{p_0 - \frac{\mathbf{p}^2}{2m} + v_F \mathbf{p}\cdot\bm{\sigma} }{\left( p_0+ i\epsilon - \frac{\mathbf{p}^2}{2m}\right)^2 - v_F^2\mathbf{p}^2 }\;,\\
\tilde{\Delta}^{A}(p) &=& i\frac{ p_0 - \frac{\mathbf{p}^2}{2m} + v_F \mathbf{p}\cdot\bm{\sigma}}{\left( p_0 - i\epsilon - \frac{\mathbf{p}^2}{2m}\right)^2 - v_F^2\mathbf{p}^2 }\;.
\label{eq_Deltas_Fourier}
\end{eqnarray}

In order to consider the finite temperature dependence of the polarization tensor, the time-domain is compactified according to
the prescription $t \rightarrow -i\tau$,
with $0 \le \tau \le \beta$, with $\beta = 1/(k_B T)$ the inverse temperature. Correspondingly, the three propagators defined above reduce to a single Euclidean one, by analytic continuation $p_0 + i\epsilon \rightarrow i p_4 + \mu$
of the retarded one.
Therefore, we define the Euclidean propagator by
\begin{eqnarray}
\tilde{\Delta}^{E}(p) = \tilde{\Delta}^{R}(p_0 + i\epsilon \rightarrow i p_4 + \mu,\mathbf{p}) =  i\frac{i p_4 + \mu - \frac{\mathbf{p}^2}{2m} + v_F \mathbf{p}\cdot\bm{\sigma} }{\left( i p_4+ \mu - \frac{\mathbf{p}^2}{2m}\right)^2 - v_F^2\mathbf{p}^2 }.
\end{eqnarray}

In particular, for the linear response theory\cite{Wen,Stefanucci,Rammer,Kamenev_011,Munoz_013,Merker_013}, we need the retarded component of the polarization tensor
\begin{eqnarray}
K_{R}^{\mu\nu}(x-y) = \int\frac{d^3 p}{(2\pi)^3} e^{i(x-y)\cdot p}\, \Pi_{R}^{\mu\nu}(p),
\end{eqnarray}
which is obtained at finite temperature from the Euclidean polarization tensor by analytic continuation
\begin{eqnarray}
\Pi_R^{kl}(\omega,\mathbf{p}) = \Pi_{E}^{kl}(i p_4 \rightarrow \omega + i\epsilon, \mathbf{p}). 
\label{eq_PiR}
\end{eqnarray}
The corresponding expression for the finite temperature, Euclidean polarization tensor is
\begin{eqnarray}
\Pi^{kl}_{E}(i p_4,\mathbf{p}) &=& \frac{e^2}{4 m^2}\frac{1}{\beta}\sum_{q_4 = \omega_n, n\in Z}  \int\frac{d^2 q}{(2\pi)^2}\Gamma_{ab}^k(p+2q)
\tilde{\Delta}_{bc}^{E}(p+q)
\Gamma_{cd}^l(p+2q)
\tilde{\Delta}_{da}^{E}(q)
\label{eq_PiE}
\end{eqnarray}
with the symbol
\begin{equation}
\Gamma_{ab}^k(p+2q)= \left[ \delta_{ab}(p + 2q)^k + 2\theta\left[\sigma^k\right]_{ab}\right],
\end{equation}
and a similar expression for $\Gamma_{cd}^l(p+2q)$. We remark that due to compactification of the time domain at finite temperature, the
component $q_4 = \omega_n$, where $\omega_n = 2\pi (n + 1/2)/\beta$ for $n \in \mathbb{Z}$ are the Fermionic Matsubara frequencies.

\section{The polarization tensor and optical conductivity}\label{vacpol}

The polarization tensor $ \Pi^{kl}(p)$ contains the information about the conductivity on the plane of this two-dimensional system and also about its light transmission properties\cite{FV-2016,Altland-Simons}.
We are interested in the consequences of the application of harmonic homogeneous electric fields which, in the temporal gauge, are related with the vector potential by $E^k=-\partial A^k/\partial t=-i \omega A^k$. Since the conductivity is determined by the linear relation between the current and the applied electric field, $J_k=\sigma_{kl} E^l$, from Eqs.\ \eqref{L10}, \eqref{Kmunu} and \eqref{eq_PiR}, we can write for the conductivity as a function of the frequency \cite{FV-2016,Altland-Simons}
\begin{equation}\label{sigma-pi}
    \sigma_{kl}(\omega)= 2\times 2 \left. \frac{\Pi_{kl}^{R}(p)}{ i \omega} \right|_{p \rightarrow (\omega, \mathbf{0})}\,,
\end{equation}
where the prefactor takes into account the valley and electronic spin degeneracy in graphene. Therefore, the real and imaginary components of the optical conductivity are given by
\begin{eqnarray}
\Re e\, \sigma_{kl}(\omega,T) = 4 \frac{\Im m\, \Pi^{R}_{kl}(\omega,T)}{\omega}
\end{eqnarray}
and
\begin{eqnarray}
\Im m\, \sigma_{kl}(\omega,T) = -4 \frac{\Re e\, \Pi^{R}_{kl}(\omega,T)}{\omega},
\end{eqnarray}
respectively. In particular, it is the real part of the conductivity tensor that determines electronic transport in the DC limit $\omega\rightarrow 0$.

In order to include finite temperature effects, we first calculate $\Pi_{kl}^{E}(\omega,\mathbf{0})$ from Eq.(\ref{eq_PiE}), and then by analytic continuation, as described in Eq.(\ref{eq_PiR}), we obtain $\Pi_{kl}^{R}(\omega,\mathbf{0})$.

The evaluation requires to calculate two integrals and an infinite sum over (Fermionic) Matsubara frequencies, as defined in Eq.(\ref{eq_PiE}).

\begin{eqnarray}
\Pi_{kl}^{E}(p) &=&\frac{e^2}{4 m^2} \frac{1}{\beta}\sum_{q_4 = \omega_n, n\in Z} \int\frac{d^2 q}{(2\pi)^2}{\rm{Tr}}\left\{\left[ p_k + 2q_k + 2\theta \sigma_k \right]\Delta^{E}(p+q)
\left[ p_l + 2q_l + 2\theta\sigma_l\right]\Delta^{E}(q)\right\}.
\end{eqnarray}
Specializing this expression to the case $p = \left(i p_4,\mathbf{0}\right)$, and using polar coordinates
for the spatial components $q_1=Q \cos\varphi, q_2=Q \sin\varphi$, we write
\begin{eqnarray}
\Pi_{kl}^{E}(i p_4,\mathbf{0}) = \frac{e^2}{4 \pi} \frac{1}{\beta}\sum_{q_4 = \omega_n, n\in Z}  \int_0^{\infty}\frac{d Q\,Q}{4\pi m^2} \int_{0}^{2\pi} d\varphi \frac{{\rm{Tr}}\{A\}}{B^{EE}}
\label{eq_PiFA1}
\end{eqnarray}
with
\begin{eqnarray}
A &=&\left[ 2q_k + 2\theta \sigma_k \right]\left[ i p_4 + i q_4 + \mu - \frac{\mathbf{q}^2}{2m} + v_F \mathbf{q}\cdot\bm{\sigma}\right]
\left[ 2q_l + 2\theta\sigma_l\right]\left[ i q_4 + \mu - \frac{\mathbf{q}^2}{2m} + v_F \mathbf{q}\cdot\bm{\sigma}\right]\;, \nonumber\\
B^{EE}&=&\left( \left(i p_4 + i q_4 + \mu - \frac{\mathbf{q}^2}{2m}\right)^2 - v_F^2 \mathbf{q}^2\right)
\left(\left( i q_4 + \mu - \frac{\mathbf{q}^2}{2m}\right)^2 - v_F^2\mathbf{q}^2\right).
\label{PiE_den}
\end{eqnarray}
We notice that the denominator is independent of $\varphi$, and hence it is straightforward to 
calculate the trace in the numerator integrated over $\varphi$,
\begin{eqnarray}\label{trazas-integ}
     N(Q,i p_4, i q_4 + \mu) &=& \frac{1}{4\pi m^2}\int_0^{2\pi} {\rm Tr} \left\{ A \right\} d\varphi\nonumber\\
      &=&   -\left(8 \left(8 m^4 v_f^2 (i q_4 + \mu ) (   i q_4 + \mu + i p_4 )
     +4 m^2 Q^2 \left(i p_4  \left(
     m v_f^2+i q_4 + \mu \right)\right.\right.\right.\nonumber\\
     &&\left.\left.\left.+(i q_4 + \mu ) \left(i q_4 + \mu +2 m v_f^2 \right)\right)-2 m Q^4 \left( m v_f^2+2 i q_4+ 2 \mu +   i p_4 \right)
     +Q^6\right)\right.
     \end{eqnarray}
for $k,l=1,1$ or $2,2$, and a vanishing result for $k,l=1,2$ or $2,1$.

\begin{figure}[htbp]
	\centering
		\includegraphics[width=0.5\textwidth]{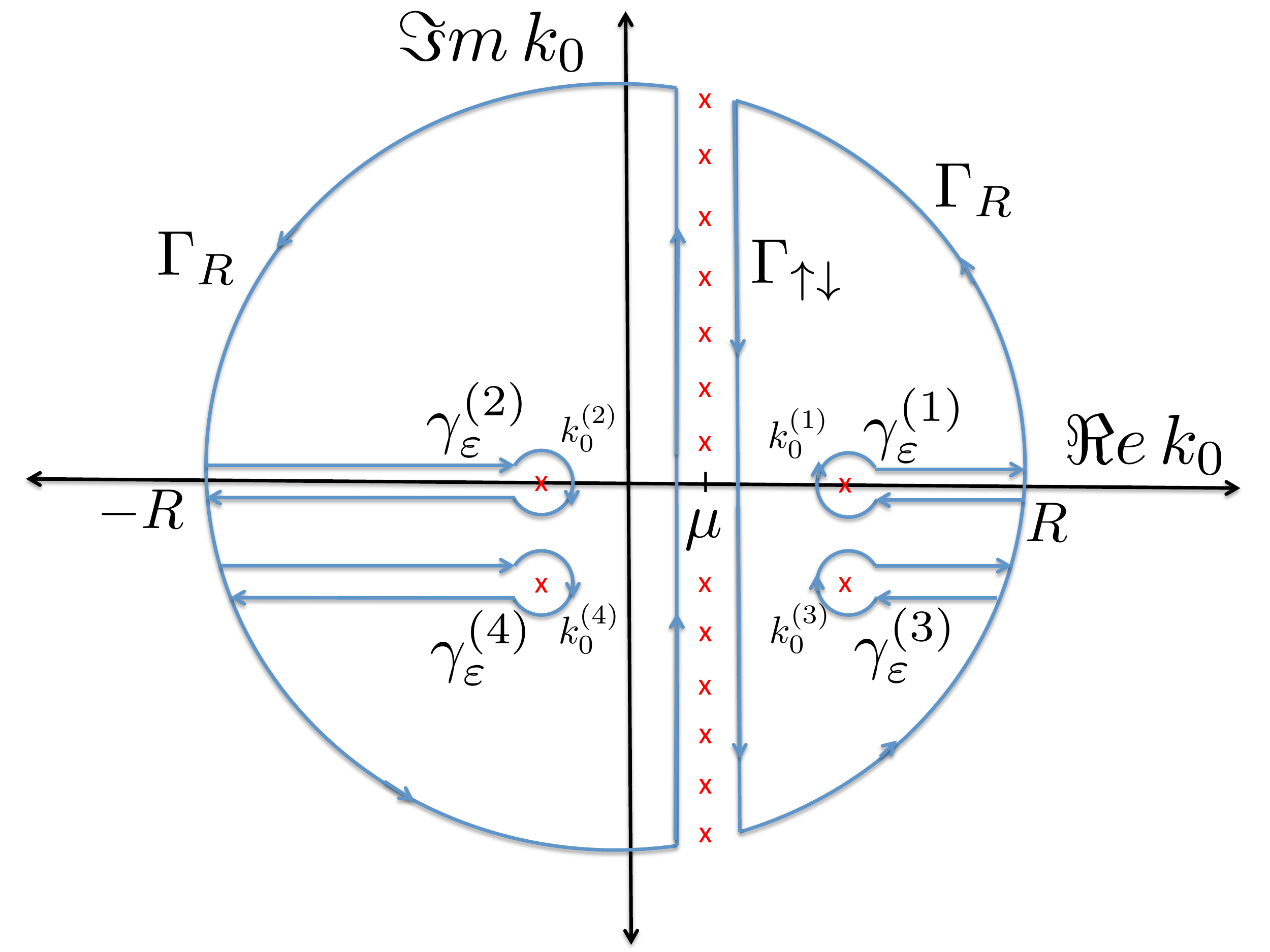}
	\caption{(Color online) The complex contour $C = \Gamma_R\oplus\Gamma_{\uparrow\downarrow}\oplus_{\alpha}\gamma_{\varepsilon}^{(\alpha)}$ used to calculate the Matsubara sum. Notice that $\Gamma_{\uparrow\downarrow}$
	and $\gamma_{\varepsilon}^{(\alpha)}$ are oriented clockwise, in order to exclude the poles from the contour $C$.}
	\label{fig2}
\end{figure}

Let us now consider the sum over (Fermionic) Matsubara frequencies, since $q_4 = \omega_n = (2 n +1)\pi/\beta$. The sum can be obtained
through the construction of a contour integral on the complex plane (see Fig.\ref{fig2}), by choosing a meromorphic function with infinitely many poles
at the Matsubara frequencies. A straightforward choice is the Fermi function,
\begin{eqnarray}
n_F(k_0 - \mu) = \frac{1}{1 + e^{\beta(k_0 - \mu)}},
\end{eqnarray}
that clearly has poles at $k_0 = i \omega_n + \mu$, for $n \in \mathbb{Z}$, with residues
\begin{eqnarray}
\text{Res}\left[ n_F(k_0-\mu)  \right]_{k_0 = i\omega_n + \mu} &=& \lim_{k_0 \rightarrow i\omega_n + \mu} \frac{(k_0 - i\omega_n -\mu)}{1 + e^{\beta(k_0 - \mu)}} \nonumber\\
&=& \lim_{k_0 \rightarrow i\omega_n + \mu} \frac{(k_0 - i\omega_n -\mu)}{1 + e^{i\beta\omega_n}e^{\beta(k_0 - i\omega_n - \mu)}} = -\frac{1}{\beta},
\label{eq_pole_nF}
\end{eqnarray}
where the identity $e^{i\beta\omega_n} = -1$, valid for fermionic Matsubara frequencies, was applied.

Therefore, defining $i q_4 + \mu \rightarrow k_0$, we calculate the contour integral depicted in Fig.\ref{fig2}, when the radius of the outer circular contour $\Gamma_R$ goes to infinity, $R\rightarrow\infty$, and
the radius of the 4 contours $\gamma_{\epsilon}^{(\alpha)}$ goes to zero, $\varepsilon \rightarrow 0$
\begin{eqnarray}
\lim_{R\rightarrow\infty,\varepsilon\rightarrow 0}\oint_{C} \frac{N(Q, i p_4, k_0)}{B^{EE}(Q,i p_4, k_0)} n_F(k_0-\mu) \frac{dk_0}{2 \pi i} &=& -\sum_{\alpha = 1, 4} \text{Res}\left[ \frac{N(Q, i p_4, k_0)}{B^{EE}(Q,i p_4, k_0)} \right]_{k_0=k_0^{(\alpha)}} n_F(k_0^{(\alpha)}-\mu)\nonumber\\
&-& \sum_{n\in Z}  \frac{N(Q, i p_4, i \omega_n + \mu)}{B^{EE}(Q,i p_4, i \omega_n + \mu)} \text{Res}\left[ n_F(k_0-\mu)  \right]_{k_0 = i\omega_n + \mu} = 0. 
\end{eqnarray}

Using Eq.(\ref{eq_pole_nF}), we solve for the required Matsubara sum from the equation above,
\begin{eqnarray}
\frac{1}{\beta} \sum_{n\in Z}  \frac{N(Q, i p_4, i \omega_n + \mu)}{B^{EE}(Q,i p_4, i \omega_n + \mu)} = 
\sum_{\alpha = 1, 4} \text{Res}\left[ \frac{N(Q, i p_4, k_0)}{B^{EE}(Q,i p_4, k_0)} \right]_{k_0^{(\alpha)}} n_F(k_0^{(\alpha)}-\mu).
\end{eqnarray}
Here, the poles are the roots of the denominator of the quartic polynomial, i.e. $B^{EE}(Q, i p_4, k_0^{(\alpha)}) = 0$, for $\alpha = 1,\ldots,4$. Explicitly, one finds
\begin{eqnarray}
k_0^{(1)} &=& \frac{Q(Q + 2 m v_f)}{2 m},\nonumber\\
k_0^{(2)} &=& \frac{Q(Q - 2 m v_f)}{2 m},\nonumber\\
k_0^{(3)} &=& \frac{Q(Q + 2 m v_f)}{2 m} - i p_4,\nonumber\\
k_0^{(4)} &=& \frac{Q(Q - 2 m v_f)}{2 m} - i p_4.
\end{eqnarray}
By recalling that the external Matsubara frequency in the diagram is a Bosonic one, we have $p_4 = 2 n \pi/\beta$, with $n\in\mathbb{Z}$, and hence $e^{i \beta p_4 } = 1$. Using this
simple  identity, we find that
\begin{eqnarray}
n_F( k_0^{(3)} - \mu) = n_F(k_0^{(1)} - \mu),\,\,\,\,\, n_F(k_0^{(4)} - \mu) = n_F(k_0^{(2)} - \mu).
\end{eqnarray}
Using this, and calculating explicitly the residues, we finally obtain
\begin{eqnarray}
\Pi_{11}^{E}(i p_4,\mathbf{0}) = \frac{e^2}{4 \pi} \int_0^{\infty} dQ\,\frac{4 v_f^3 Q^2}{4 v_f^2 Q^2 - (i\,p_4)^2} \left( n_F\left[\frac{Q(Q - 2m v_f)}{2m}- \mu 
\right] - n_F\left[\frac{Q(Q + 2m v_f)}{2m} - \mu 
\right] \right)
\label{eq_PiE_int} 
\end{eqnarray}

From this expression, by analytic continuation to real frequency space $i p_4 \rightarrow \omega + i\epsilon$ we recover the retarded polarization tensor
\begin{eqnarray}
\Pi_{11}^{R}(\omega) = \Pi_{11}^{E}(\mathbf{0}, i p_4 \rightarrow \omega + i\epsilon).
\end{eqnarray}
For this purpose, we write part of the integrand in Eq.~(\ref{eq_PiE_int}) as follows
\begin{eqnarray}
\frac{4 v_f^3 Q^2}{4 v_f^2 Q^2 - (\omega + i\epsilon)^2} &=& v_f^2 Q \left[
\frac{1}{2 v_f Q - \omega - i\epsilon} + \frac{1}{2 v_f Q + \omega + i\epsilon}
\right]\nonumber\\
&=& \mathcal{P} \frac{4 v_f^3 Q^2}{4 v_f^2 Q^2 - \omega^2} + i \pi v_f^2 Q \left[
\delta(2 v_f Q - \omega) - \delta(2 v_f Q + \omega)
\right],
\end{eqnarray}
where $\mathcal{P}$ stands for the Cauchy principal value. Therefore, the real and imaginary parts of the retarded polarization tensor
are given by the expressions

\begin{eqnarray}
\Re e\,\Pi_{11}^{R}(\omega) = \frac{e^2}{4 \pi} \mathcal{P} \int_0^{\infty} dQ\,\frac{4 v_f^3 Q^2}{4 v_f^2 Q^2 - \omega^2} \left( n_F\left[\frac{Q(Q - 2m v_f)}{2m}- \mu 
\right] - n_F\left[\frac{Q(Q + 2m v_f)}{2m} - \mu 
\right] \right) 
\end{eqnarray}

\begin{eqnarray}
\Im m\, \Pi_{11}^{R}(\omega) &=& \frac{e^2}{4} v_f^2 \int_{0}^{\infty} dQ\,Q\left[
\delta(2 v_f Q - \omega) - \delta(2 v_f Q + \omega)
\right] \left( 
n_F\left[\frac{Q(Q - 2m v_f)}{2m}- \mu 
\right]\right.\nonumber\\
&&\left. - n_F\left[\frac{Q(Q + 2m v_f)}{2m} - \mu 
\right]
\right)
\end{eqnarray}

Moreover, in order to remove unphysical, possibly divergent vacuum contributions from the retarded polarization tensor, we define its regularized version as
\begin{eqnarray}
\Pi_{11,\,reg}^{R}(\omega) \equiv \Pi_{11}^{R}(\omega,T) - \Pi_{11}^{R}(0,T). 
\end{eqnarray}
Note from the definitions above that, by construction, $\Im m\, \Pi_{11}^{R}(\omega = 0, T) = 0$, and hence no regularization is required for the imaginary part of the tensor. On the other hand, $\Re e\, \Pi_{11}(\omega = 0,T) \ne 0$ in general, and hence the real part will be regularized as described in Appendix.
The expression for the real part cannot be reduced to a simple analytical expression, however one can still evaluate it in a low-temperature series through
a generalization of Sommerfeld expansion (as shown in Appendix). On the other hand, the integral for the imaginary part can be evaluated to yield
\begin{eqnarray}
\Im m\, \Pi_{11}^{R}(\omega) &=& \frac{e^2}{16}\omega\sign(\omega)\left(
n_F\left[ \frac{\omega^2}{8 m v_f^2} - \frac{\omega}{2} - \mu \right] - n_F\left[ \frac{\omega^2}{8 m v_f^2} + \frac{\omega}{2} - \mu \right]
\right)\nonumber\\
&=& \frac{e^2}{32}|\omega|\left(
\tanh\left[\frac{\beta}{2}\left( \frac{\omega^2}{8 m v_f^2} + \frac{\omega}{2} - \mu\right) \right] - \tanh\left[\frac{\beta}{2}\left( \frac{\omega^2}{8 m v_f^2} - \frac{\omega}{2} - \mu \right)\right]
\right).
\label{eq_Imsigma_T}
\end{eqnarray}
From the expression above, the real part of the optical conductivity is given by
\begin{eqnarray}
\Re e\, \sigma_{11}(\omega,T) &=& 4 \frac{\Im m\, \Pi_{11}^{R}(\omega)}{\omega}\nonumber\\
&=& \frac{e^2}{8\hbar}\sign(\omega)\left(
\tanh\left[\frac{\beta}{2}\left( \frac{\hbar^2\omega^2}{8 m v_f^2} + \frac{\hbar\omega}{2} - \mu\right) \right] - \tanh\left[\frac{\beta}{2}\left( \frac{\hbar^2\omega^2}{8 m v_f^2} - \frac{\hbar\omega}{2} - \mu \right)\right]
\right),
\label{eq_Resigma_T}
\end{eqnarray}
where we have restored the $\hbar$ constant for normal I.S. units.

\begin{figure}
	\centering
	\subfloat[$t' = 0.056\,eV$]{
		\includegraphics[width=0.5\textwidth]{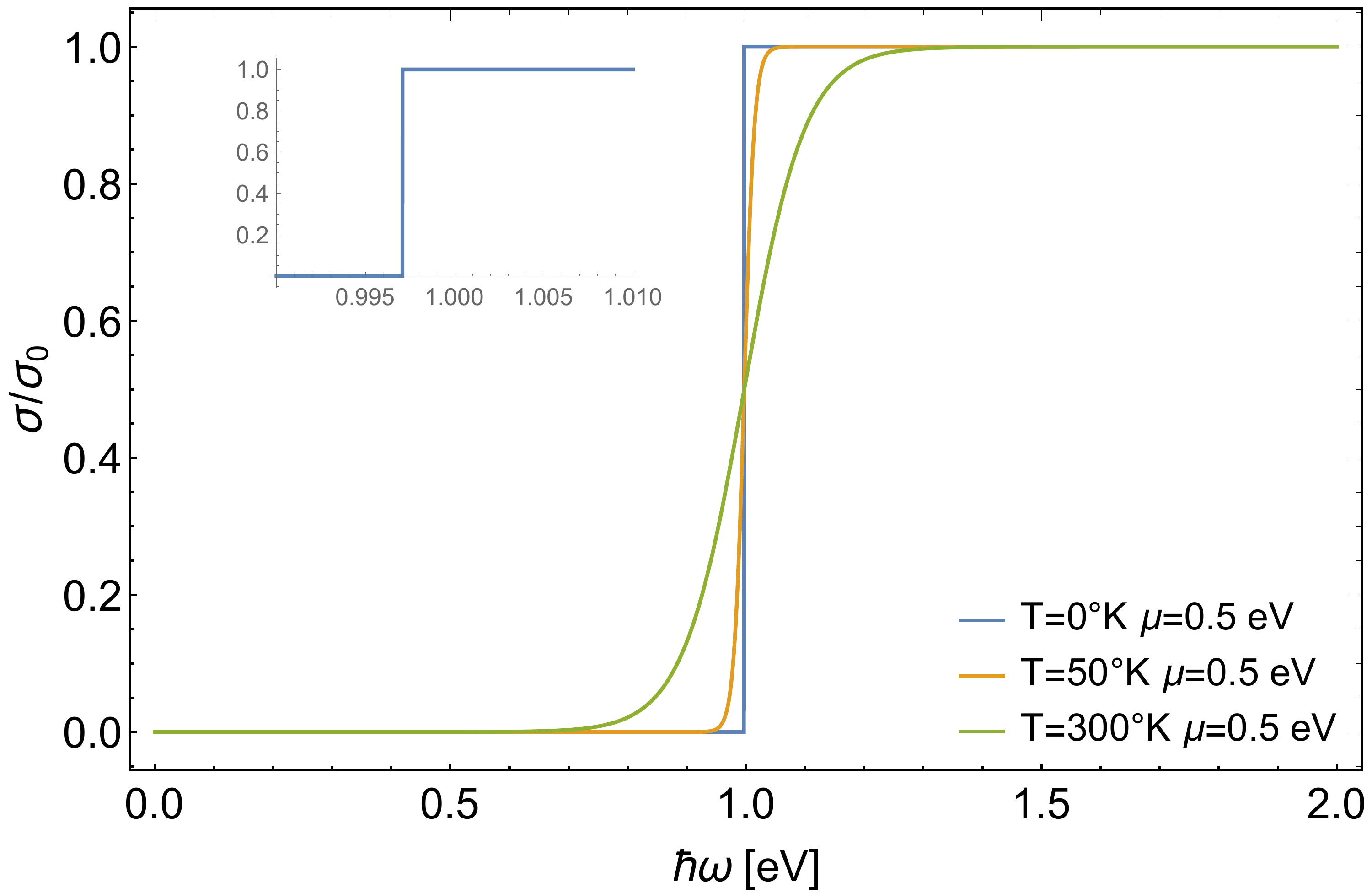}
		\label{fig3a}
		}
	\subfloat[$t' = 0.56\,eV$]{
	\includegraphics[width=0.5\textwidth]{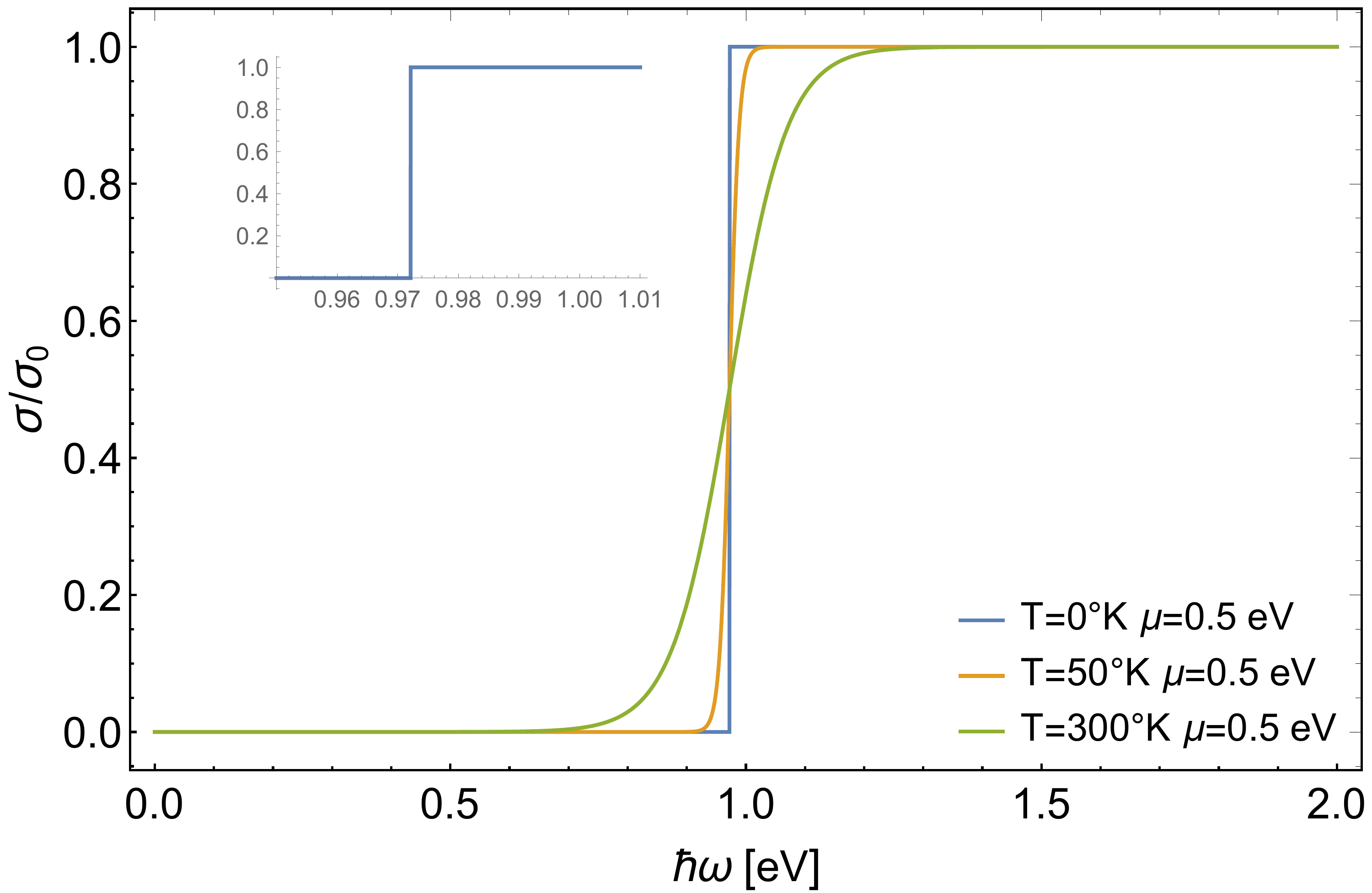}	
	\label{fig3b}
	}
	\caption{(Color online) The real part of the electrical conductance, for (a) $t' = 0.056\,eV$, and (b) $t' = 0.56\,eV$ (see Table~\ref{tab:table1}), at constant chemical potential $\mu = 0.5\,eV$, as a function of frequency, for different temperature vales.}
\end{figure}

It is very interesting to analyze the zero-temperature limit ($\beta\rightarrow\infty$) of Eq.(\ref{eq_Resigma_T}),
that becomes (see Appendix B for details)
\begin{eqnarray}
\Re e\, \sigma_{11}(\omega,T \rightarrow 0) = \left\{
\begin{array}{cc}
 \frac{e^2}{4\hbar}, & \sqrt{1 + \frac{2 \mu}{m v_f^2}} - 1 < \frac{\hbar|\omega|}{2 m v_f^2} <   \sqrt{1 + \frac{2 \mu}{m v_f^2}} + 1\\
 0, & \text{otherwise}
\end{array}
\right.
\label{eq_Resigma_0}
\end{eqnarray}
It is seen from this result that the actual value of the conductivity at $T = 0$ is $e^2/(4\hbar)$, independent of frequency and the parameter $m$ that
captures the second nearest-neighbor interaction, in agreement with our previous calculation~\cite{Cond_T0} and transparency experiments~\cite{Nair}. Interestingly though, there is however a hidden, non-analytic dependency through the domain of the stepwise
function, that defines a region where the conductivity actually vanishes. It is instructive to compare our result, that includes the second nearest-neighbor
interaction through the parameter $m$, with the more standard result that only involves first nearest-neighbors, a situation that can be recovered
from our model in the limit $m\rightarrow\infty$. In this limit, from Eq.~(\ref{eq_Resigma_T}) we obtain
\begin{eqnarray}
\Re e\, \sigma_{11}(\omega,T,m\rightarrow\infty) 
= \frac{e^2}{8\hbar}\sign(\omega)\left(
\tanh\left[\frac{\beta}{2}\left(  \frac{\hbar\omega}{2} - \mu\right) \right] + \tanh\left[\frac{\beta}{2}\left(  \frac{\hbar\omega}{2} + \mu \right)\right]
\right).
\label{eq_Resigma_T_m_infty}
\end{eqnarray}
This result, as expected, matches the one reported in Refs.\cite{Kuzmenko_PRL_2008,Falkovsky_PRB_2007}. Moreover, also in the limit $m\rightarrow\infty$, the zero-temperature conductivity
becomes
\begin{eqnarray}
\Re e\, \sigma_{11}(\omega,0,m\rightarrow\infty) &=&  \frac{e^2}{8\hbar}\sign(\omega)\left\{\sign\left(  \hbar\omega - 2\mu\right)+ \sign\left(  \hbar\omega +2 \mu \right)
\right\}\\
& = & \left\{\begin{array}{cc}0\,, \quad |\omega|<2\mu/\hbar \\
\frac{e^2}{4\hbar} \,, \quad |\omega|>2\mu/\hbar.\end{array}\right.
\label{eq_Resigma_T0_m_infty}
\end{eqnarray}
in agreement with Refs.\cite{Nair,Kuzmenko_PRL_2008}. The real part of the electrical conductance, as a function of frequency and at different temperatures, is depicted in Fig.~\ref{fig3a} and Fig.~\ref{fig3b}.

Let us now turn to the imaginary part of the optical conductivity. The integral over $0 \le Q < \infty$ can be expressed as an asymptotic expansion in negative powers of $\beta$, through
a similar analysis as in the more standard Sommerfeld expansion (for details see Appendix). The real part of the retarded polarization tensor (see Appendix) is given by the expression
\begin{eqnarray}
\Re e\,\Pi_{11,reg}^{R}(\omega,T)&=& 
\frac{e^2}{8 \pi}\omega
\mathcal{F}(\omega,\mu,m) + \beta^{-2}\frac{e^2 \pi \omega^2}{24 m v_f^2 \left(1 + \frac{2\mu}{m v_f^2} \right)^{3/2}}
 \left(
 \frac{\omega^2 - 8 m v_f^2\left(3 \mu + 2 m v_f^2\left(1 + \sqrt{1 + \frac{2\mu}{m v_f^2}} \right) \right)}{\left[\omega^2 - 8 m v_f^2\left( \mu +  m v_f^2\left(1 + \sqrt{1 + \frac{2\mu}{m v_f^2}} \right) \right)\right]^2}
 \right.\nonumber\\
 &&\left.- \frac{\omega^2 + 8 m v_f^2\left(-3 \mu + 2 m v_f^2\left(-1 + \sqrt{1 + \frac{2\mu}{m v_f^2}} \right) \right)}{\left[\omega^2 - 8 m v_f^2\left( \mu +  m v_f^2\left(-1 + \sqrt{1 + \frac{2\mu}{m v_f^2}} \right) \right)\right]^2}
 \Theta\left[\frac{\mu}{m v_f^2}\right]
 \right) + O(\beta^{-3}).
\end{eqnarray}
Therefore, the imaginary part of the optical conductivity is given by
\begin{eqnarray}
\Im m\, \sigma_{11}(\omega) &=& -4 \frac{\Re e\,\Pi_{11,reg}^{R}(\omega,T)}{\omega}\nonumber\\
&=& -\frac{e^2}{2 \pi \hbar}\mathcal{F}(\omega,\mu,m) 
 - (k_B T)^{2}\frac{e^2 \pi \omega}{6 m v_f^2 \left(1 + \frac{2\mu}{m v_f^2} \right)^{3/2}}
 \left(
 \frac{\hbar^2\omega^2 - 8 m v_f^2\left(3 \mu + 2 m v_f^2\left(1 + \sqrt{1 + \frac{2\mu}{m v_f^2}} \right) \right)}{\left[\hbar^2\omega^2 - 8 m v_f^2\left( \mu +  m v_f^2\left(1 + \sqrt{1 + \frac{2\mu}{m v_f^2}} \right) \right)\right]^2}
 \right.\nonumber\\
 &&\left.- \frac{\hbar^2\omega^2 + 8 m v_f^2\left(-3 \mu + 2 m v_f^2\left(-1 + \sqrt{1 + \frac{2\mu}{m v_f^2}} \right) \right)}{\left[\hbar^2\omega^2 - 8 m v_f^2\left( \mu +  m v_f^2\left(-1 + \sqrt{1 + \frac{2\mu}{m v_f^2}} \right) \right)\right]^2}
 \Theta\left[\frac{\mu}{m v_f^2}\right]
 \right) + O(\beta^{-3}),
 \label{Imag_sigma}
\end{eqnarray}
where we have restored the $\hbar$ constant for I.S. units, and we defined the function
\begin{eqnarray}
\mathcal{F}(\omega,\mu,m) = \left\{
 \begin{array}{cc}
  {\rm{arctanh}}\left[\frac{\hbar\omega}{2 m v_f^2\left(\sqrt{1 + \frac{2\mu}{m v_f^2}} - 1\right)} \right] - 
  {\rm{arctanh}}\left[\frac{\hbar\omega}{2 m v_f^2\left(\sqrt{1 + \frac{2\mu}{m v_f^2}} + 1\right)} \right] ,& 0 < \hbar\omega < 2 m v_f^2\left(\sqrt{1 + \frac{2 \mu}{m v_f^2}} -1\right)\nonumber\\
 \frac{1}{2} \ln\left[\frac{\left(\sqrt{1 + \frac{2\mu}{m v_f^2}} + 1 - \frac{\hbar\omega}{2 m v_f^2}\right)}{\left(\sqrt{1 + \frac{2\mu}{m v_f^2}} + 1 + \frac{\hbar\omega}{2 m v_f^2}\right)} \frac{\left(\sqrt{1 + \frac{2\mu}{m v_f^2}} - 1 + \frac{\hbar\omega}{2 m v_f^2}\right)}{\left(\frac{\hbar\omega}{2 m v_f^2}-\sqrt{1 + \frac{2\mu}{m v_f^2}} + 1\right) } \right],& \sqrt{1 + \frac{2 \mu}{m v_f^2}} -1 < \frac{\hbar\omega}{2 m v_f^2} < \sqrt{1 + \frac{2 \mu}{m v_f^2}} +1\nonumber\\
  {\rm{arctanh}}\left[\frac{2 m v_f^2\left(\sqrt{1 + \frac{2\mu}{m v_f^2}} - 1\right)}{\hbar\omega} \right] - 
  {\rm{arctanh}}\left[\frac{2 m v_f^2\left(\sqrt{1 + \frac{2\mu}{m v_f^2}} + 1\right)}{\hbar\omega} \right],&\hbar\omega > 2 m v_f^2\left(\sqrt{1 + \frac{2 \mu}{m v_f^2}} +1\right).
 \end{array}
 \right.
 \end{eqnarray}

\begin{figure}
	\centering
	\subfloat[$t' = 0.056\,eV$]{
		\includegraphics[width=0.5\textwidth]{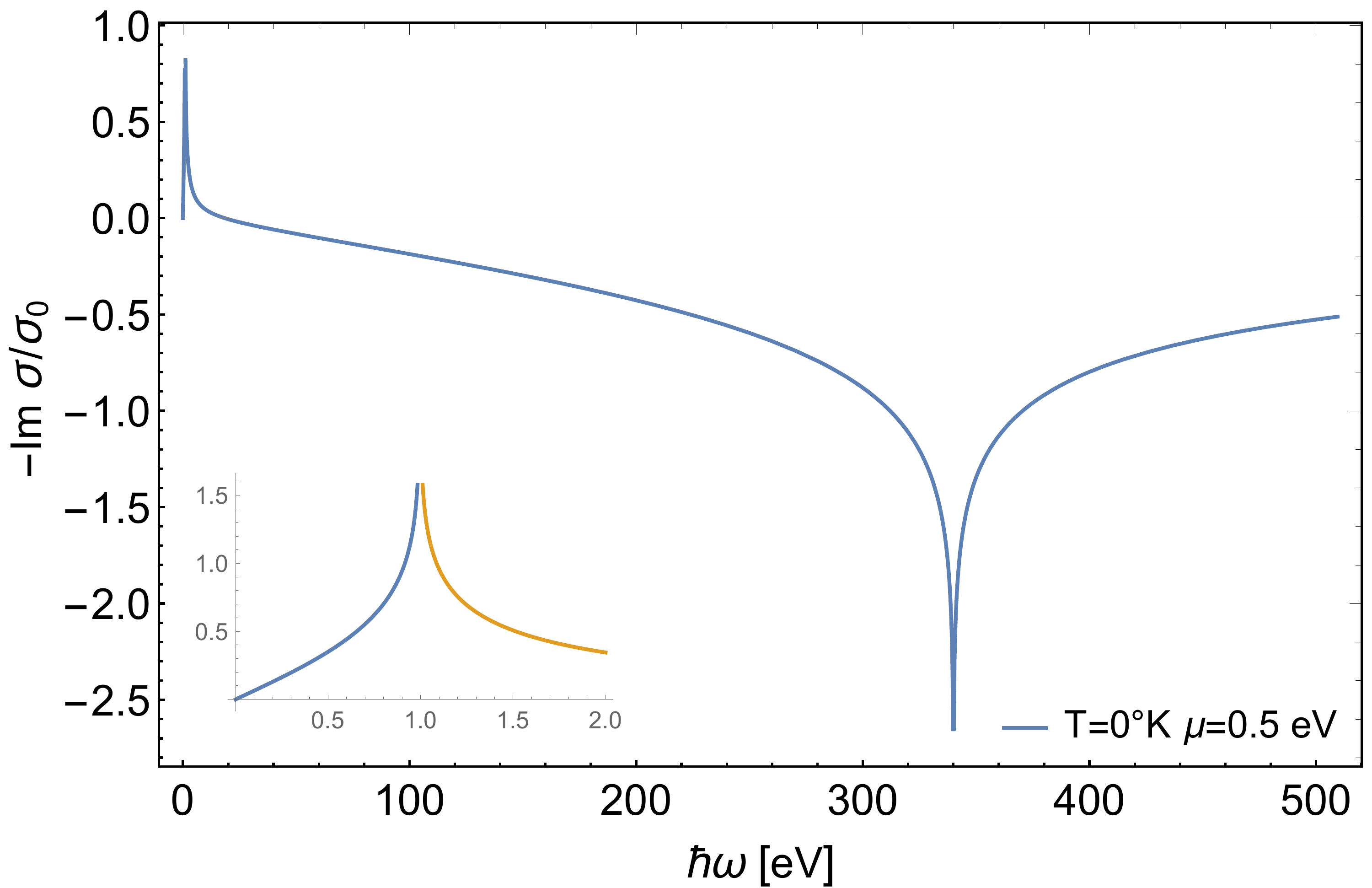}
		\label{fig4a}
		}
	\subfloat[$t' = 0.56\,eV$]{
	\includegraphics[width=0.5\textwidth]{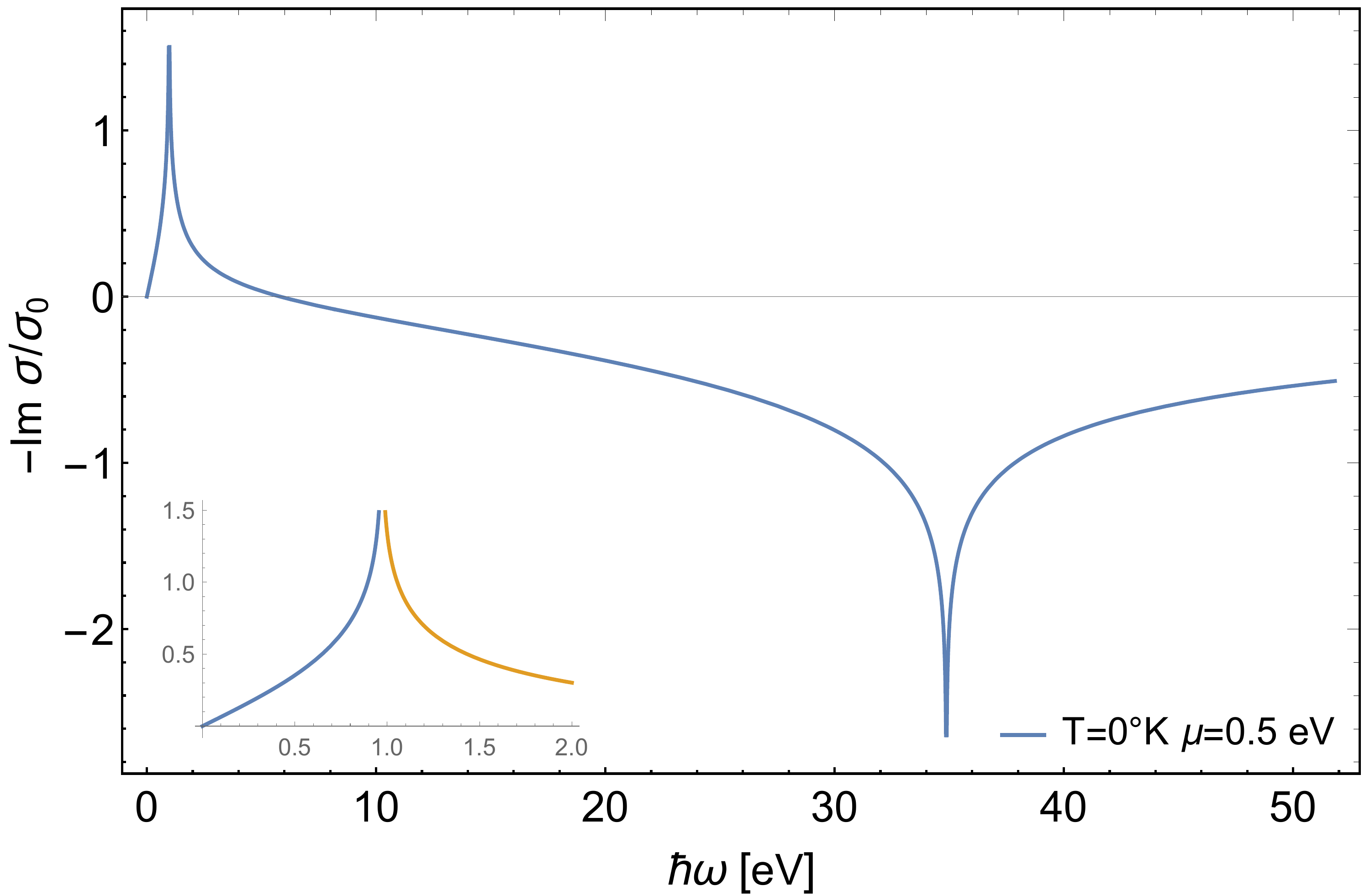}	
	\label{fig4b}
	}
	\caption{(Color online) The Imaginary part of the electrical conductance, for (a) $t' = 0.056\,eV$, and (b) $t' = 0.56\,eV$ (see Table~\ref{tab:table1}), at constant chemical potential $\mu = 0.5\,eV$, as a function of frequency, at
	zero temperature. The finite temperature dependence is very weak (as seen in Eq.(\ref{Imag_sigma})
	and cannot be appreciated at the scale of the plot. The inset shows with higher resolution the region near the first peak.}
\end{figure}

The imaginary part of the optical conductivity, expressed in our model by Eq.(~\ref{Imag_sigma}), displays two separate resonances (see Fig.~\ref{fig4a} and Fig.~\ref{fig4b}), the first at $\hbar\omega = 2 m v_f^2\left( \sqrt{1 + \frac{2 \mu}{m v_f^2}} - 1  \right)\sim 2\mu - \frac{\mu^2}{m v_f^2}$, and the second at $\hbar\omega = 2 m v_f^2\left( \sqrt{1 + \frac{2 \mu}{m v_f^2}} + 1  \right)\sim 2 m v_f^2$. 
The first one reproduces, in the limit $m\rightarrow\infty$, results reported in the literature for the conventional model with only first-to-nearest neighbor approximation \cite{Nair,Kuzmenko_PRL_2008}, with a small shift $\sim - \frac{\mu^2}{m v_f^2}$ in the position of the peak. The second peak, which is a unique feature of the model, is located at
an extremely large frequency, and in practice has no physical consequences.

\section{Conclusions}\label{conclusions}

Along this article, we have discussed the effect of including the next-to-nearest neighbors hopping $t'$, through the "mass" parameter $m = \pm 2 \hbar^2/(9 t' a^2)$ in the dispersion relation \cite{GNAQ}, on the optical conductivity of
single-layer graphene. Our analysis is based on the continuum representation of the model via an effective field theory\cite{Cond_T0}, by extending our previous results at zero temperature \cite{Cond_T0}
to the finite chemical potential and finte temperature scenario, Eq.(\ref{eq_Resigma_T}) and Eq.(~\ref{Imag_sigma}). As expected, our analytical calculation recovers the universal value $\Re e\,\sigma = e^2/(4\hbar)$ in the limit of zero temperature, Eq.(\ref{eq_Resigma_0}), but however reveals a non-trivial
and non-analytic dependence on the ratio $\mu/(m v_f^2)$ in the frequency domain. Remarkably, our analytical Eq.(\ref{eq_Resigma_T}) for the frequency-dependent real part of the optical conductivity at finite temperature and chemical potential, in the limit $m\rightarrow\infty$ ($t'\rightarrow 0$) reduces to Eq.(\ref{eq_Resigma_T_m_infty}), 
that exactly reproduces previous results reported in the literature \cite{Nair,Kuzmenko_PRL_2008} for the conventional first-nearest-neighbor approximation. Moreover, our Eq.(\ref{eq_Resigma_T}) generalizes this result to reveal the effect of including the next-to-nearest neighbor hopping $t'$ into the dispersion relation. In particular,
we notice that, when $t'$ is neglected as in the conventional case, the real part of the conductivity presents a sharp step (at zero temperature) or a sigmoidal trend (at finite temperature) exactly centered at $\hbar\omega = 2\mu$ (see for instance Eq.(\ref{eq_Resigma_T_m_infty})). In contrast,
when $t'$ is included, the step is shifted to $\hbar\omega = 2 m v_f^2\left( \sqrt{1 + \frac{2 \mu}{m v_f^2}} - 1  \right)\sim 2\mu - \frac{\mu^2}{m v_f^2}$. This effect is particularly interesting since, as shown in the existing literature, there seems to be a large uncertainty on the exact value for the second nearest neighbor hopping
in graphene, $0.056\,eV\, < t' <\, 0.56\,eV$ (see Table~\ref{tab:table1}). Our result suggests that an experimental characterization of the frequency-dependence of the real part of the optical conductivity, at finite chemical potential (to be adjusted, for instance, with a gate potential) could therefore provide an accurate and direct 
experimental measurement of $t'$, that could be compared with the broad estimations obtained so far from ab-initio calculations \cite{Reich_2002} or cyclotron resonance experiments \cite{Deacon_2007}.

\section*{Acknowledgements}
H.F. thanks ANPCyT, CONICET and UNLP, Argentina, for partial support through grants PICT-2014-2304, PIP 2015-688 and Proy. Nro. 11/X748, respectively. H.F. also acknowledges PUC for its kind hospitality. E. M. acknowledges support from FONDECYT (Chile) under grant No. 1190361. M. Loewe acknowledges support from FONDECYT (Chile) under grants No. 1170107 and No. 1190192.  R. Zamora would like to thank support from CONICYT FONDECYT Iniciaci\'on under grant No. 11160234. 

\appendix

\section{Zero temperature limit of $\Re e\, \sigma_{11}(\omega,T)$}

Let us start from Eq.(\ref{eq_Resigma_T}) (in natural units $\hbar=1$), and consider the limit $T \rightarrow 0$ ($\beta \rightarrow \infty$), 
\begin{eqnarray}
\Re e\, \sigma_{11}(\omega,T = 0) 
= \frac{e^2}{8}\sign(\omega)\left(
\sign\left[\frac{\omega^2}{4 m v_f^2} + \omega - 2 \mu \right] - \sign\left[\frac{\omega^2}{4 m v_f^2} - \omega - 2\mu \right]
\right).
\label{eq_Resigma_App}
\end{eqnarray}

\begin{figure}
	\centering
		\includegraphics[width=0.5\textwidth]{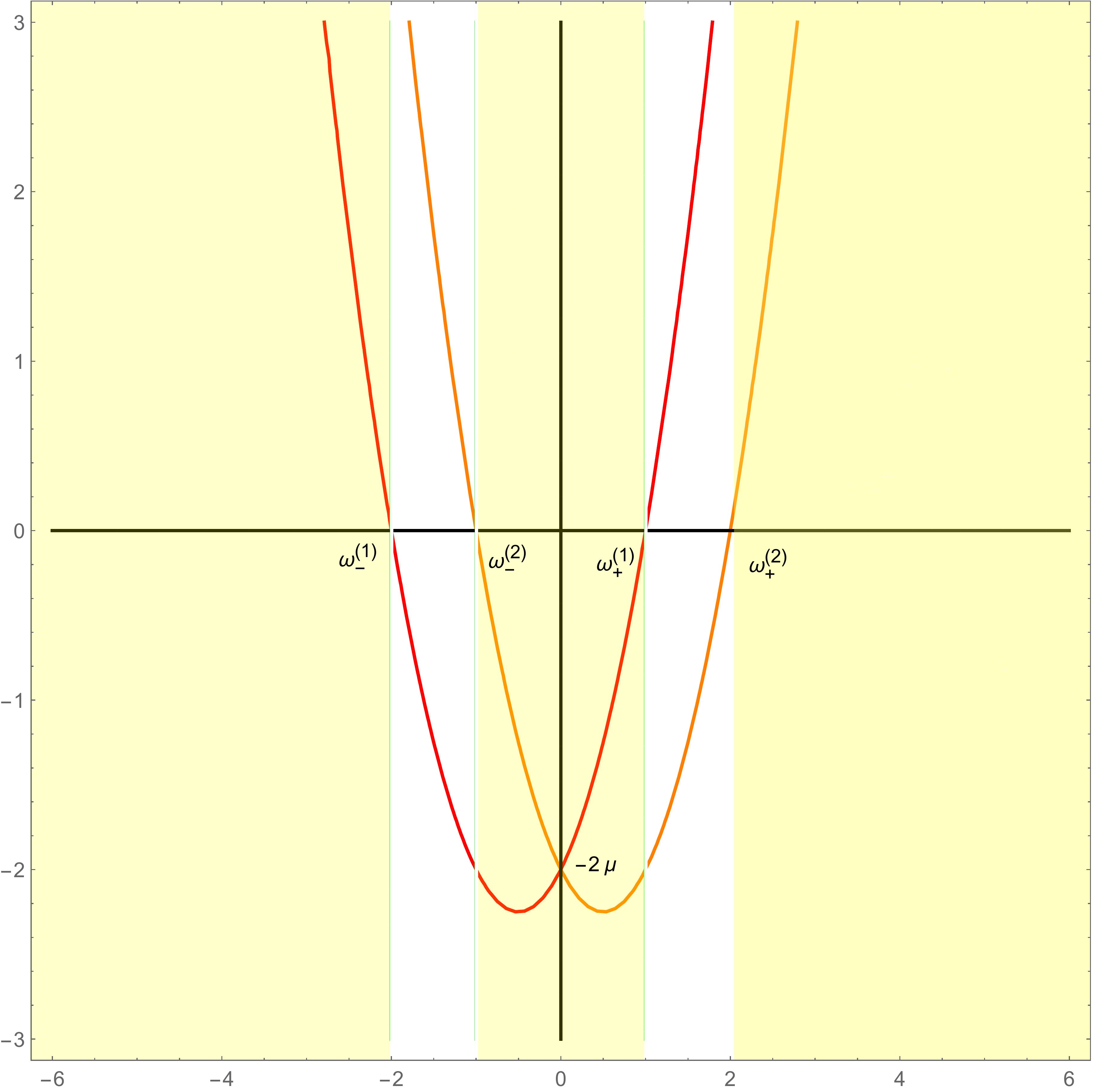}
	\caption{(Color online) Sketch of the locus of the roots in Eq.~(\ref{eq_roots}). The regions in white represent the frequency range where, at zero temperature and finite chemical potential, the real part of the optical conductivity
	does not vanish, as seen in Eq.~(\ref{Real_cond_T0_App}).}
	\label{fig5}
\end{figure}

Clearly, the difference between the $\sign(z)$ functions is either $\pm 2$ or $0$. In order to analyze the different cases, let us define the two quadratic functions
\begin{eqnarray}
y_1(\omega) &=& \frac{\omega^2}{4 m v_f^2} + \omega - 2 \mu = (\omega - \omega^{(1)}_{+})(\omega - \omega^{(1)}_{-}),\nonumber\\
y_2(\omega) &=& \frac{\omega^2}{4 m v_f^2} - \omega - 2 \mu = (\omega - \omega^{(2)}_{+})(\omega - \omega^{(2)}_{-}),
\end{eqnarray}
where roots are given by
\begin{eqnarray}
\omega_{\pm}^{(1)} &=& -2 m v_f^2 \pm 2 m v_f^2 \sqrt{1 + \frac{2 \mu}{m v_f^2}},\nonumber\\
\omega_{\pm}^{(2)} &=& 2 m v_f^2 \pm 2 m v_f^2 \sqrt{1 + \frac{2 \mu}{m v_f^2}}.
\label{eq_roots}
\end{eqnarray}
On the other hand, the two parabolas intersect at $\omega = 0$,  with the common value $y_1(0) = y_2(0) = -2\mu$. A graphical representation
of the roots and intercept is displayed in Fig.~\ref{fig5}. Moreover, we remark that Eq.(\ref{eq_Resigma_App}) can be written as
\begin{eqnarray}
\Re e\, \sigma_{11}(\omega,T = 0) 
= \frac{e^2}{8}\sign(\omega)\left(
\sign(y_1) - \sign(y_2)
\right) &=& \frac{e^2}{4}\left\{\begin{array}{cc}
\sign(\omega), & y_1(\omega) > 0, \,\, y_2(\omega) < 0\\
-\sign(\omega), & y_1(\omega) < 0, \,\, y_2(\omega) > 0\\
0, & \text{otherwise}
\end{array}
\right.
\end{eqnarray}

The condition $y_1(\omega) > 0$ and  $y_2(\omega) < 0$ is satisfied for $-2 m v_f^2 + 2 m v_f^2 \sqrt{1 + \frac{2 \mu}{m v_f^2}} < \omega < 2 m v_f^2 + 2 m v_f^2 \sqrt{1 + \frac{2 \mu}{m v_f^2}}$,
where $\sign(\omega) = 1$. On the other hand, the condition $y_1(\omega) < 0$ and  $y_2(\omega) > 0$ is satisfied for $-2 m v_f^2 - 2 m v_f^2 \sqrt{1 + \frac{2 \mu}{m v_f^2}} < \omega < 2 m v_f^2 - 2 m v_f^2 \sqrt{1 + \frac{2 \mu}{m v_f^2}}$, where $\sign(\omega) = -1$. Taking this into account, we arrive at the final expression
\begin{eqnarray}
\Re e\, \sigma_{11}(\omega,T \rightarrow 0) = \left\{
\begin{array}{cc}
 \frac{e^2}{4\hbar}, & \sqrt{1 + \frac{2 \mu}{m v_f^2}} - 1 < \frac{\hbar|\omega|}{2 m v_f^2} <   \sqrt{1 + \frac{2 \mu}{m v_f^2}} + 1\\
 0, & \text{otherwise}
\end{array}
\right..
\label{Real_cond_T0_App}
\end{eqnarray}
where we have restored the $\hbar$ constant for I.S. units.

\section{Low temperature expansion for $\Re e\, \Pi_{11}^{R}(\omega)$}

Let us consider the integral representing the real part of the retarded polarization tensor
\begin{eqnarray}
\Re e\,\Pi_{11}^{R}(\omega) = \frac{e^2}{4 \pi} \mathcal{P} \int_0^{\infty} dQ\,\frac{4 v_f^3 Q^2}{4 v_f^2 Q^2 - \omega^2} \left( n_F\left[\frac{Q(Q - 2m v_f)}{2m}- \mu 
\right] - n_F\left[\frac{Q(Q + 2m v_f)}{2m} - \mu 
\right] \right), 
\end{eqnarray}
where $\mathcal{P}$ stands for Cauchy's principal value.

It is convenient to express the integral defining the polarization tensor in dimensionless variables, i.e.
\begin{eqnarray}
x = Q/(m v_f),\,\,\,\, \Omega = \omega/(2 m v_f^2), \,\,\,\, \bar{\beta} = m v_f^2 \beta / 2, \,\,\,\, \gamma = 2\mu/(m v_f^2).
\end{eqnarray}
Hence, we have
\begin{eqnarray}
\Re e\,\Pi_{11}^{R}(\omega) = \frac{e^2}{4 \pi}m v_f^2  \mathcal{P}  \int_0^{\infty} dx\, \frac{x^2}{x^2 - \Omega^2}\,\left[ \bar{n}_F(x^2 - 2x- \gamma) 
 - \bar{n}_F(x^2 + 2x- \gamma) \right], 
\end{eqnarray}
with the Fermi distributions at the dimensionless $\bar{\beta}$,
\begin{eqnarray}
\bar{n}_F(z) = \left(1 + e^{\bar{\beta}z}\right)^{-1}.
\end{eqnarray}

As discussed in the main text, in order to remove spurious unphysical and possibly divergent contributions arising from the vacuum, we regularize the
retarded polarization tensor according to the expression
\begin{eqnarray}
\Re e\,\Pi_{11,reg}^{R}(\omega,T) &\equiv& \Re e\,\Pi_{11}^{R}(\omega,T) - \Re e\,\Pi_{11}^{R}(0,T).
\end{eqnarray}
It is interesting first to analyze the $T\rightarrow 0$ limit of the regularized polarization tensor. From the expression for the Fermi functions,
it is clear that $\bar{n}_F(z) \rightarrow \Theta(-z)$ as $\bar{\beta} \rightarrow \infty$ ($T\rightarrow 0$). Therefore, we have
\begin{eqnarray}
\Re e\,\Pi_{11,reg}^{R}(\omega,T\rightarrow 0) &=&  \frac{e^2}{4 \pi}m v_f^2 \Omega^2  \mathcal{P}  \int_0^{\infty} dx\, \frac{1}{x^2 - \Omega^2}\,\left[ \Theta(x^2 + 2x- \gamma) 
 - \Theta(x^2 - 2x- \gamma) \right],\nonumber\\
 &=& \frac{e^2}{4 \pi}m v_f^2 \Omega^2  \mathcal{P} \int_{x_{+}^{(1)}}^{x_{+}^{(2)}}\frac{dx}{x^2 - \Omega^2},
 \end{eqnarray}
 where $x_{+}^{(1)} = \sqrt{1 + \gamma} -1$ and $x_{+}^{(2)} = \sqrt{1 + \gamma} +1$ are the positive roots of the quadratic polynomials $y_1(x) = x^2 + 2x- \gamma$
and $y_2(x) = x^2 - 2x- \gamma$, respectively.
 The principal value integral must be calculated separately in three frequency intervals, giving the results
 \begin{eqnarray}
 \mathcal{P} \int_{x_{+}^{(1)}}^{x_{+}^{(2)}}\frac{dx}{x^2 - \Omega^2} = \left\{
 \begin{array}{cc}
 \frac{1}{\Omega}\left[ {\rm{arctanh}}(\Omega/x_{+}^{(1)}) -  {\rm{arctanh}}(\Omega/x_{+}^{(2)})\right],& 0 < \Omega < x_{+}^{(1)}\\
 \frac{1}{2\Omega} \ln\left[\frac{x_{+}^{(2)} - \Omega}{x_{+}^{(2)}+\Omega} \frac{x_{+}^{(1)} + \Omega}{\Omega - x_{+}^{(1)}} \right],& x_{+}^{(1)} < \Omega < x_{+}^{(2)}\\
  \frac{1}{\Omega}\left[ {\rm{arctanh}}(x_{+}^{(1)}/\Omega) -  {\rm{arctanh}}(x_{+}^{(2)}/\Omega)\right],&\Omega > x_{+}^{(2)}
 \end{array}
 \right.
 \end{eqnarray}

 Therefore, we have the analytical expression
\begin{eqnarray}
\Re e\,\Pi_{11,reg}^{R}(\omega,T\rightarrow 0) = \frac{e^2}{8 \pi}\omega\mathcal{F}(\omega,\mu,m)
\end{eqnarray}
where we have defined the function
\begin{eqnarray}
\mathcal{F}(\omega,\mu,m) = \left\{
 \begin{array}{cc}
  {\rm{arctanh}}\left[\frac{\omega}{2 m v_f^2\left(\sqrt{1 + \frac{2\mu}{m v_f^2}} - 1\right)} \right] - 
  {\rm{arctanh}}\left[\frac{\omega}{2 m v_f^2\left(\sqrt{1 + \frac{2\mu}{m v_f^2}} + 1\right)} \right] ,& 0 < \omega < 2 m v_f^2\left(\sqrt{1 + \frac{2 \mu}{m v_f^2}} -1\right)\nonumber\\
 \frac{1}{2} \ln\left[\frac{\left(\sqrt{1 + \frac{2\mu}{m v_f^2}} + 1 - \frac{\omega}{2 m v_f^2}\right)}{\left(\sqrt{1 + \frac{2\mu}{m v_f^2}} + 1 + \frac{\omega}{2 m v_f^2}\right)} \frac{\left(\sqrt{1 + \frac{2\mu}{m v_f^2}} - 1 + \frac{\omega}{2 m v_f^2}\right)}{\left(\frac{\omega}{2 m v_f^2}-\sqrt{1 + \frac{2\mu}{m v_f^2}} + 1\right) } \right],& \sqrt{1 + \frac{2 \mu}{m v_f^2}} -1 < \frac{\omega}{2 m v_f^2} < \sqrt{1 + \frac{2 \mu}{m v_f^2}} +1\nonumber\\
  {\rm{arctanh}}\left[\frac{2 m v_f^2\left(\sqrt{1 + \frac{2\mu}{m v_f^2}} - 1\right)}{\omega} \right] - 
  {\rm{arctanh}}\left[\frac{2 m v_f^2\left(\sqrt{1 + \frac{2\mu}{m v_f^2}} + 1\right)}{\omega} \right],&\omega > 2 m v_f^2\left(\sqrt{1 + \frac{2 \mu}{m v_f^2}} +1\right).
 \end{array}
 \right.
 \end{eqnarray}

For the finite temperature contribution, we obtain
\begin{eqnarray}
\Re e\,\Pi_{11,reg}^{R}(\omega,T)= \Re e\,\Pi_{11,reg}^{R}(\omega,T\rightarrow 0) + \frac{e^2}{2 \pi}m v_f^2 \left( \Pi_1(\omega) - \Pi_2(\omega) - \Pi_1(0) + \Pi_2(0)  \right),
\end{eqnarray}
where
\begin{eqnarray}
\Pi_1(\omega) &=& 2\sum _{k=0}^N  \beta ^{-2 k-2}\left(1-2^{-2 k-1}\right) \zeta (2 k+2) F_{+}^{(2k+1)}(0)
\nonumber\\
&+&\delta _{\gamma }\left[\sum _{k=1}^{2 N} \beta ^{-k-1} \left(1-2^{-k}\right) (-1)^k \zeta (k+1) F_{-}^{(k)}(0)+ \beta^{-1} F_{-}(0) \log (2)\right]_{\gamma\rightarrow 0}.
\end{eqnarray}
\begin{eqnarray}
\Pi_2(\omega) &=& 2 \theta (\gamma ) \sum _{k=0}^N \beta ^{-2 k-2} \left(1-2^{-2 k-1}\right) \zeta (2 k+2) G_{+}^{(2 k+1)}(0)\nonumber\\
&+&\delta _{\gamma } \left[\sum _{k=1}^{2 N} \beta ^{-k-1}\left(1-2^{-k}\right) (-1)^k \zeta (k+1) G_{+}^{(k)}(0)+ \beta^{-1} G_{+}(0) \log (2)\right]_{\gamma\rightarrow 0}
\end{eqnarray}

In these expressions, we have defined the auxiliary functions obtained from the roots of the quadratic equations $x^2 \pm 2 x - \gamma = z$, corresponding to
\begin{eqnarray}
x^{(1)}_{\pm}(z) &=& 1 \pm \sqrt{1 + \gamma + z},\nonumber\\
x^{(2)}_{\pm}(z) &=& -1 \pm  \sqrt{1 + \gamma + z},
\end{eqnarray}
and the corresponding implicit functions
\begin{eqnarray}
F_{\pm}(z) &=& \frac{f[x_{\pm}^{(1)}(z)]}{2(\, x_{\pm}^{(1)}(z) - 1)}\nonumber\\
G_{\pm}(z) &=& \frac{f[x_{\pm}^{(2)}(z)]}{2(\, x_{\pm}^{(2)}(z) + 1)},
\end{eqnarray}
where we defined the function
\begin{eqnarray}
f(x) = \frac{x^2}{x^2 -\Omega^2},
\end{eqnarray}

Similarly, in the above expansions we defined the derivatives of these implicit functions with respect to $z$, as
\begin{eqnarray}
F^{(k)}_{\pm}(0) = \left.\frac{d^{k}}{dz^{k}} F_{\pm}(z)\right|_{z = 0},\,\,\,\,
G^{(k)}_{\pm}(0) = \left.\frac{d^{k}}{dz^{k}} G_{\pm}(z)\right|_{z = 0}.
\end{eqnarray}

The explicit expression
for finite temperature corrections up to $O(\beta^{-3})$ is
\begin{eqnarray}
\Re e\,\Pi_{11,reg}^{R}(\omega,T)&=& 
\frac{e^2}{8 \pi}\omega
\mathcal{F}(\omega,\mu,m) + \beta^{-2}\frac{e^2 \pi \omega^2}{24 m v_f^2 \left(1 + \frac{2\mu}{m v_f^2} \right)^{3/2}}
 \left(
 \frac{\omega^2 - 8 m v_f^2\left(3 \mu + 2 m v_f^2\left(1 + \sqrt{1 + \frac{2\mu}{m v_f^2}} \right) \right)}{\left[\omega^2 - 8 m v_f^2\left( \mu +  m v_f^2\left(1 + \sqrt{1 + \frac{2\mu}{m v_f^2}} \right) \right)\right]^2}
 \right.\nonumber\\
 &&\left.- \frac{\omega^2 + 8 m v_f^2\left(-3 \mu + 2 m v_f^2\left(-1 + \sqrt{1 + \frac{2\mu}{m v_f^2}} \right) \right)}{\left[\omega^2 - 8 m v_f^2\left( \mu +  m v_f^2\left(-1 + \sqrt{1 + \frac{2\mu}{m v_f^2}} \right) \right)\right]^2}
 \Theta\left[\frac{\mu}{m v_f^2}\right]
 \right) + O(\beta^{-3})
\end{eqnarray}

Here, we have defined the Heaviside Theta function as
\begin{eqnarray}
\theta(x) = \left\{\begin{array}{cc}
1, & x > 0\\
0, & x \le 0.
\end{array}\right.
\end{eqnarray}


\end{document}